\documentclass[aps,pre,onecolumn,showpacs,square,numbers,amssymb,amsmath,superscriptaddress,nobibnotes,nofootinbib]{revtex4-1}
\usepackage{bm}%
\usepackage[colorlinks=true,linkcolor=red]{hyperref}
\usepackage{amsmath}
\usepackage{amsfonts}
\usepackage{amssymb}
\usepackage{verbatim}
\usepackage{epsfig}
\usepackage{graphicx}
\usepackage[sort&compress]{natbib}
\usepackage{tikz}
\usetikzlibrary{snakes}

\def\cp{\mathcal{P}}
\begin{document}


\title{Joint probability densities of level spacing ratios in random matrices}


\author{Y.~Y.~Atas}
\author{E.~Bogomolny}
\author{O.~Giraud}
\author{P.~Vivo}
\affiliation{Univ.~Paris Sud, CNRS, LPTMS, UMR8626, Orsay F-91405 (France)}
\author{E.~Vivo}
\affiliation{Departamento de Matem\'aticas and Grupo Interdisciplinar de Sistemas Complejos (GISC), Universidad Carlos III de Madrid, Avenida de la Universidad 30, E-28911 Legan\'es (Spain)}


\date{May 30, 2013}

\begin{abstract}
We calculate analytically, for finite-size matrices, joint probability densities of ratios of level spacings in ensembles of random matrices characterized by their associated confining potential. 
We focus on the ratios of two spacings between three consecutive real eigenvalues, as well as certain generalizations such as the overlapping ratios. The resulting formulas are further analyzed in detail in two specific cases: the $\beta$-Hermite and the $\beta$-Laguerre cases, for which we offer explicit calculations for small $N$. The analytical results are in excellent agreement with numerical simulations of usual random matrix ensembles, and with the level statistics of a quantum many-body lattice model and zeros of the Riemann zeta function.
\end{abstract}


\maketitle


\section{Introduction}

Since its inception more than fifty years ago, Random Matrix Theory (RMT) has experienced successful applications in a variety of fields \cite{mehta,oxford,reviewRMT,for}. The original idea was to replace a sufficiently complicated Hermitian operator (the Hamiltonian of heavy nuclei) with a big matrix filled with random numbers and then to study the statistical regularities of the energy spectra. Later, in the field of quantum chaos it was rapidly realized that the statistics of the \emph{spacings} between consecutive levels could help discriminating between systems whose classical counterpart is chaotic or integrable \cite{berry,bohigas}. However, different random matrix ensembles or Hamiltonian systems may (and usually do) have different mean level densities, and a meaningful comparison between spacing distributions requires a transformation called \emph{unfolding}. The unfolded levels $\tilde{\lambda}_i$ and the true levels $\lambda_i$ are related via $\tilde{\lambda}_i = \mathcal{N}(\lambda_i)$, where $\mathcal{N}(x)$ is the mean number of levels less than $x$. The unfolded spectrum has mean level spacing equal to $1$ and facilitates comparison between different models, provided that $\mathcal{N}(x)$ is analytically known or easily estimated.

However, in many-body problems a reliable determination of $\mathcal{N}(x)$ is often computationally difficult. In order to overcome the necessity of unfolding, a new spectral measure was proposed in \cite{OgaHus07}. Let $\{\lambda_i\}$ be a set of ordered energy levels and $s_n = \lambda_{n+1}-\lambda_n$ be the nearest-neighbor spacings. Oganesyan and Huse \cite{OgaHus07} studied the statistics of the ratios
\begin{equation}
\label{defratios}
\tilde{r}_n=\min\left(r_n,\frac{1}{r_n}\right),\qquad r_n = \frac{s_{n+1}}{s_{n}}\, .
\end{equation}
The use of ratios of consecutive spacing makes the unfolding procedure unnecessary, as such quantities are expected to be independent of the local density of states. Numerical studies of related quantities have recently appeared in the literature on finite size lattices \cite{Kollath10,Santos10,Collura12} and on many-body localization \cite{OgaHus07,Kollath10,Santos10,Collura12,ioffe}. While gap probabilities in matrix models are by now fairly well understood (see e.g. \cite{for,gap1,gap2} and references therein), analytical works on ratios of consecutive spacings were virtually non-existent until a very recent paper \cite{atas} where a `Wigner-like' surmise was proposed for the probability density of ratios \eqref{defratios}, based on an exact calculation for $3\times 3$ Gaussian random matrices. The surmise was shown to provide an excellent approximation for the probability density of the ratios of arbitrarily large matrices and the remaining tiny difference could be fitted by a simple interpolating formula. Applications were given to a quantum many-body Hamiltonian and to the zeros of the Riemann zeta function, in all cases with excellent agreement between the numerics and the analytical surmise.

Before summarizing our contribution to these problems, we recall that a very general class of so-called \emph{$\beta$-ensembles} of $N\times N$ random matrices is defined in terms of the joint probability density of the $N$ real eigenvalues as
\begin{equation}
P_\beta^{(V)}(\lambda_1,\ldots,\lambda_N)=\frac{1}{Z_{N,\beta}^{(V)}}e^{-\sum_{i=1}^N V(\lambda_i)}\prod_{j<k}|\lambda_j-\lambda_k|^\beta\,,
\label{P_beta}
\end{equation}
where $Z_{N,\beta}^{(V)}$ is a normalization constant. Here $V(x)$ is a confining potential suitably growing at infinity and $\beta$ is the Dyson index of the ensemble, which can take any real positive value. Here we focus on so-called $\beta$-Hermite and $\beta$-Laguerre ensembles, characterized respectively by $V(x)=x^2/2$ and $V(x)=x/2-(\alpha-1) \ln x$. Explicit matrix realizations of these ensembles for any $\beta>0$ are possible thanks to the work by Dumitriu and Edelman \cite{dumitriu}. The resulting matrices are tridiagonal, with independent but not identically distributed entries. For the special cases $\beta=1,2,4$ non-sparse matrix representations are also available. Namely, the $\beta$-Hermite ensemble includes real symmetric ($\beta=1$), complex Hermitian ($\beta=2$) or quaternion self-dual ($\beta=4$) $N\times N$ matrices with independent Gaussian distributed entries, respectively denoted GOE, GUE and GSE ensembles. The eigenvalues are real. The $\beta$-Laguerre ensemble includes $N\times N$ so-called Wishart (covariance) matrices of the form $\mathbf{W}=\mathbf{X}^\dagger \mathbf{X}$ with $\mathbf{X}$ a rectangular $M\times N$ real ($\beta=1$), complex ($\beta=2$) or quaternionic ($\beta=4$) matrix with independent Gaussian distributed entries. The eigenvalues of $\mathbf{W}$ are real and non-negative. We refer to \cite{maj} for an excellent review on Wishart ensembles.

The aim of the present paper is to consider various generalizations of the results obtained in \cite{atas}. As mentioned above, the analytical expression obtained in \cite{atas} for the distribution of ratios of adjacent spacings, based on an explicit calculation for the $3\times 3$ $\beta$-Hermite ensemble, was shown to approximate well the exact Fredholm determinant formula, valid for $N\to\infty$. However, as was observed numerically in \cite{atas}, the $4\times 4$ distribution is a much better approximation to the large-$N$ result than the $3\times 3$ distribution (at least for $\beta=2$), but no analytic expression for the $N=4$ case yet exists. Moreover, it is highly desirable to go beyond an approximate evaluation (however accurate), and possibly to extend the range of applicability to a wider class of matrix models, and also to other types of ratios. The purpose of this paper is thus threefold:
\begin{enumerate}
\item We extend the calculations of \cite{atas} by analytically deriving the ratio distribution in the case of $N=4$ eigenvalues. We thus obtain an explicit analytic expression that is more accurate by an order of magnitude than the expression of \cite{atas}.
\item We generalize the previous results, providing an \emph{exact} general formula (expressed in terms of a double-integral) for the joint probability density $\cp_\beta^{(V)} (r_1,\ldots, r_{N-2})$ of the ratios $r_j$ of consecutive spacings $r_j = (\lambda_{j+2}-\lambda_{j+1})/(\lambda_{j+1}-\lambda_{j})$, valid for \emph{any} $\beta$-ensemble of random matrices characterized by the potential $V(x)$. The general formula \eqref{general} given below is then specialized to the $\beta$-Hermite (formula \eqref{finalgaussian}) and $\beta$-Laguerre (formula \eqref{finalwishart}) cases and is valid for any $N\geq 3$ and $\beta>0$. It is more conveniently expressed in terms of auxiliary variables. The surmise in \cite{atas} then becomes a special case of the general formula \eqref{finalgaussian} for $N=3$, and we give extensions to the cases $N=4,5$ computing the one-point marginals $\rho_{\beta,N}^{(V)}(r)$ of the joint density. These specific examples for small $N$, once worked out explicitly, hint towards an interesting \emph{universal} behavior of the marginal densities, namely 
\begin{align}
\rho_{\beta,N}^{(V)}(r) &\sim r^\beta \qquad\mbox{for } r\to 0\nonumber\\
\rho_{\beta,N}^{(V)}(r) &\sim r^{-2-\beta} \quad\mbox{for } r\to\infty,\label{asbeh}
\end{align}
 \emph{independently} of the confining potential $V(x)$ and $N$. In contrast, it is known that this universality does \emph{not} hold for the spacing surmises (see \cite{lag3} for a detailed discussion). 
\item We consider another kind of generalization, namely the $k$th overlapping ratio $(\lambda_{n+k+1}-\lambda_n)/(\lambda_{n+k}-\lambda_{n-1})$. We provide analytical expressions for Poisson distribution, and as an illustrative case for  $\beta$-Hermite ensemble of random matrices in the case $k=1$. These results are then applied to spectral properties of a quantum Ising model and to zeros of the Riemann zeta function.
\end{enumerate}

The plan of the paper is as follows. In Section \ref{ratios34} we recall the derivation of the ratio distribution for $N=3$ and extend it to the more accurate $N=4$ case. We then generalize these calculations to arbitrary $N$ in Section \ref{Jointpro} , and apply this to the $\beta$-Hermite case (subsection \ref{betahermite}) and to the $\beta$-Laguerre case (subsection \ref{betalaguerre}). We then turn to the $k$th overlapping ratio distributions in Section \ref{sec.overlap}. Finally, concluding remarks are offered in Section \ref{conclusions}.

\section{Ratios and spacing distributions for small matrix size}
\label{ratios34}

\subsection{Poisson and semi-Poisson distributions}
For independent and uniformly distributed random variables, the distribution of the ratio $r_n=s_{n+1}/s_{n}$ is readily calculated from the nearest-neighbour distribution $P_0(s)=\exp(-s)$, and reads $\cp_0(r)=1/(1+r)^2$. A similar calculation can be easily performed in the case of the so-called semi-Poisson distribution \cite{semipoiss}, where eigenvalues are characterized by their nearest-neighbour distribution $P_1(s)=4s\exp(-2s)$. This calculation can be generalized to distributions with exponential decrease of the nearest-neighbor spacing distribution and level repulsion as $s^{\nu}$. Their general expression is 
\begin{equation}
P_{\nu}(s)=\frac{\Gamma(\nu+2)^{\nu+1}}{\Gamma(\nu+1)^{\nu+2}}s^{\nu}\exp\left(-\frac{\Gamma(\nu+2)}{\Gamma(\nu+1)}s\right)\label{poiss}, 
\end{equation}
where the constants are chosen so that the distribution is normalized and $\langle s\rangle=1$. The semi-Poisson distribution corresponds to the case $\nu=1$. The corresponding ratio distribution is easily calculated from \eqref{poiss} and yields
\begin{equation}
\label{poisson_nu}
\cp_{\nu}(r)=\frac{\Gamma(2\nu+2)\Gamma^2(\nu+2)}{(\nu+1)^2\Gamma^4(\nu+1)}\frac{r^\nu}{(1+r)^{2\nu+2}}.
\end{equation}
For $\nu=0$ we recover the Poisson case.

\subsection{Small-size matrices: $N=3$ and $N=4$}\label{small34}
The analytical expression obtained in \cite{atas} for the distribution of ratios of adjacent spacings was derived by considering a $3\times 3$ $\beta$-Hermite ensemble and computing explicitly the marginal density (see \eqref{densitydef}) $\rho_{\beta,3}^{(\mathrm{H})}(r)$ of the ratio of the two consecutive spacings between the three eigenvalues as
\begin{equation}
\label{pwr3}
\rho_{\beta,3}^{(\mathrm{H})}(r)\propto \int_{-\infty}^\infty d\lambda_2\int_{-\infty}^{\lambda_2}d\lambda_1\int_{\lambda_2}^\infty d\lambda_3 P_\beta^{(\mathrm{H})}(\lambda_1,\lambda_2,\lambda_3)\delta\left(r-\frac{\lambda_3-\lambda_2}{\lambda_2-\lambda_1}\right),
\end{equation}
where $P_\beta^{(\mathrm{H})}(\lambda_1,\lambda_2,\lambda_3)$ is the joint density of three ordered eigenvalues of the $\beta$-Hermite ensemble, given by \eqref{P_beta} with $V(x)=x^2/2$. The integrations can be carried out explicitly and the final result reads
\begin{equation}
\rho_{\beta,3}^{(\mathrm{H})}(r)=\frac{1}{Z_\beta}\frac{(r+r^2)^{\beta}}{(1+r+r^2)^{1+\frac{3}{2}\beta}},\label{surmise3}
\end{equation}
where $Z_{\beta}$ is a proportionality constant given by
\begin{equation}
Z_\beta =\frac{2\pi\Gamma(1+\beta)}{3^{3(1+\beta)/2}\Gamma(1+\beta/2)^2}\,.
\end{equation}

 Translational invariance in the spectrum implies a left-right symmetry in the joint density of the two spacings, ultimately resulting in the duality relation
\begin{equation}
\rho_{\beta,3}^{(\mathrm{H})}(r)=\frac{1}{r^2}\rho_{\beta,3}^{(\mathrm{H})}\left(\frac{1}{r}\right).
\label{duality3}
\end{equation}
In a similar way, it is quite easy to derive the same distribution for $4\times 4$ matrices. In this case, there are two ratios $(\lambda_4-\lambda_3)/(\lambda_3-\lambda_2)$ and $(\lambda_3-\lambda_2)/(\lambda_2-\lambda_1)$. The distribution for the first ratio can be expressed similarly as in Eq.~\eqref{pwr3}. It involves the following integral
\begin{equation}
g(r)=\int_{-\infty}^\infty d\lambda_2\int_{-\infty}^{\lambda_2}d\lambda_1\int_{\lambda_2}^\infty d\lambda_3\int_{\lambda_3}^\infty d\lambda_4 P_\beta^{(\mathrm{H})}(\lambda_1,\lambda_2,\lambda_3,\lambda_4)\delta\left(r-\frac{\lambda_4-\lambda_3}{\lambda_3-\lambda_2}\right). 
\end{equation}
Changing variables $s_{i}=\lambda_{i+1}-\lambda_i$ for $i=3,2,1$, and $\lambda_2=x$, we get
\begin{eqnarray}
\label{pwr4}
g(r)=\int_{-\infty}^\infty\hspace{-.3cm} dx\int_{0}^{\infty}\hspace{-.3cm}ds_1ds_2ds_3 \left[
s_3(s_2+s_3)(s_1+s_2+s_3)s_2(s_1+s_2)s_1\right]^{\beta}s_2\delta(rs_2-s_3)\\
\times \exp\left[-\frac12\left((x-s_1)^2+x^2+(x+s_2)^2+(x+s_2+s_3)^2\right)\right].
\nonumber
\end{eqnarray}
The integral over $s_3$ can be trivially performed, and the integral over $x$ just yields an overall constant factor, leaving the integral
\begin{equation}
g(r)\propto r^{\beta}(r+1)^{\beta}\int_{0}^{\infty}ds_1ds_2s_1^{\beta}s_2^{3\beta+1}(s_1+s_2)^{\beta}[s_1+(r+1)s_2]^{\beta}\exp\left[-\frac{3}{8}s_1^2-\frac{3r^2+4r+4}8s_2^2-\frac{r+2}{4}s_1s_2\right].
\end{equation}
The remaining integrals can be performed analytically. The distribution of the second ratio $(\lambda_3-\lambda_2)/(\lambda_2-\lambda_1)$ can be obtained in the same way, yielding the whole ratio distribution. The general analytical formulae will be given in Section \ref{calculHn4}. Here, we provide the final expression for $\beta=2$ as an illustration. It reads
\begin{equation}
 \label{pderN4-I}
\rho_{2,4}^{(\mathrm{H})}(r)=\frac{1}{4\pi}\left[f(r)+\frac{1}{r^2}f\left(\frac{1}{r}\right)\right],
\end{equation}
where $f$ is the function 
\begin{equation}
f(r)=\frac{r^2 (r + 1)^2}{(1 + r + r^2)^7 (4 + 4 r + 3 r^2)^{9/2}}\left[
-(r+2)Q_1(r) + 9\sqrt{3} (4 + 4 r + 3 r^2)^4  \sqrt{4 + 4 r + 3 r^2}Q_2(r)\right]
\end{equation}
with polynomials $Q_1$ and $Q_2$ given by 
\begin{eqnarray}
\label{pderN4-II}
Q_1(r)&=&41664 + 291648 r + 946144 r^2 + 1885440 r^3 
+  2588464 r^4 + 2610064 r^5 + 2182624 r^6 + 1894048 r^7 \nonumber\\
&+&  1973866 r^8 + 2026558 r^9 + 1687399 r^{10} + 1037676 r^{11} + 
     449635 r^{12} + 124362 r^{13} + 17766 r^{14}, \nonumber\\
Q_2(r)&=& 14 + 42 r + 39 r^2 + 8 r^3 + 39 r^4 +   42 r^5 + 14 r^6.
\end{eqnarray}
The result \eqref{pderN4-I}--\eqref{pderN4-II} is an improvement over the $3\times 3$ result \eqref{pwr3}. Indeed, as has been observed numerically in \cite{atas} for $\beta=2$, the density for $4\times 4$ matrices is much closer to the large-$N$ density than the density for $3\times 3$ matrices. Namely, the absolute error $|\rho_{\beta,N}^{(\mathrm{H})}(r)-\rho_{\beta,\infty}^{(\mathrm{H})}(r)|$ is of order $10^{-2}$ for $N=3$ and $10^{-3}$ for $N=4$. In fact, the density for $N\times N$ matrices gets further away from the asymptotic result when $N$ increases up to $N\simeq 8$ and then converges back to the asymptotic result, the precision of $N=4$ being reached again only for $N\gtrsim 100$ (see Fig.~3 in \cite{atas}). Thus, the $N=4$ result is the closest to the asymptotic result among all small-size matrices, implying that Eqs.~\eqref{pderN4-I}--\eqref{pderN4-II} achieve the best possible approximation for large-$N$ distribution of spacing ratios. The accuracy of Eqs.~\eqref{pderN4-I}--\eqref{pderN4-II} will be illustrated in Fig.~\ref{densityN4fig} in Section \ref{betahermite}, where the above results will be recovered from a more general approach.

Interestingly, a similar property can be investigated at the level of the nearest-neighbor spacing distribution $P_\beta(s)$, where the usual Wigner surmise corresponding to the exact $2\times 2$ result can be improved by an order of magnitude by considering the exact $3\times 3$ calculation. Before turning back to the issue of the joint density of the ratios, we now briefly discuss this result.

\subsection{Analogy with nearest-neighbor distribution}\label{nearest}
It is well known that the nearest-neighbor distribution for the $\beta$-Hermite ensembles can be well approximated by the so-called Wigner surmise 
\begin{equation}
P_{\beta}(s)=A_{\beta}s^{\beta}\mathrm{e}^{-B_{\beta}s^2}
\label{anzatz}
\end{equation}
where $A_{\beta}$ and $B_{\beta}$ are constants determined from the normalization conditions
\begin{equation}
\int_0^{\infty}P_{\beta}(s)ds=1,\qquad \int_0^{\infty}s P_{\beta}(s)ds=1.
\label{norm}
\end{equation}
For $\beta=1,2,4$ these constants read as follows:
\begin{equation}
A_1=\frac{\pi}{2},\; B_1=\frac{\pi}{4};\qquad A_2=\frac{32}{\pi^2},\; B_2=\frac{4}{\pi};\qquad A_4=\frac{2^{18}}{3^6\pi^3},\; B_4=\frac{64}{9\pi}.
\end{equation}
Expression \eqref{anzatz} can be obtained by considering the joint distribution \eqref{P_beta} of eigenvalues of $2\times 2$ random matrices with $V(\lambda)\propto \lambda^2$, calculating
\begin{equation}
p_{\beta}(s)=\int \delta\big (s-(\lambda_2-\lambda_1)\big )P_{\beta}^{(V)}(\lambda_1,\lambda_2)\, d\lambda_1 d\lambda_2,
\label{2_2} 
\end{equation}
and normalizing the answer to obey \eqref{norm}.

\begin{figure}[!t]
\centering \includegraphics*[width=0.8\linewidth]{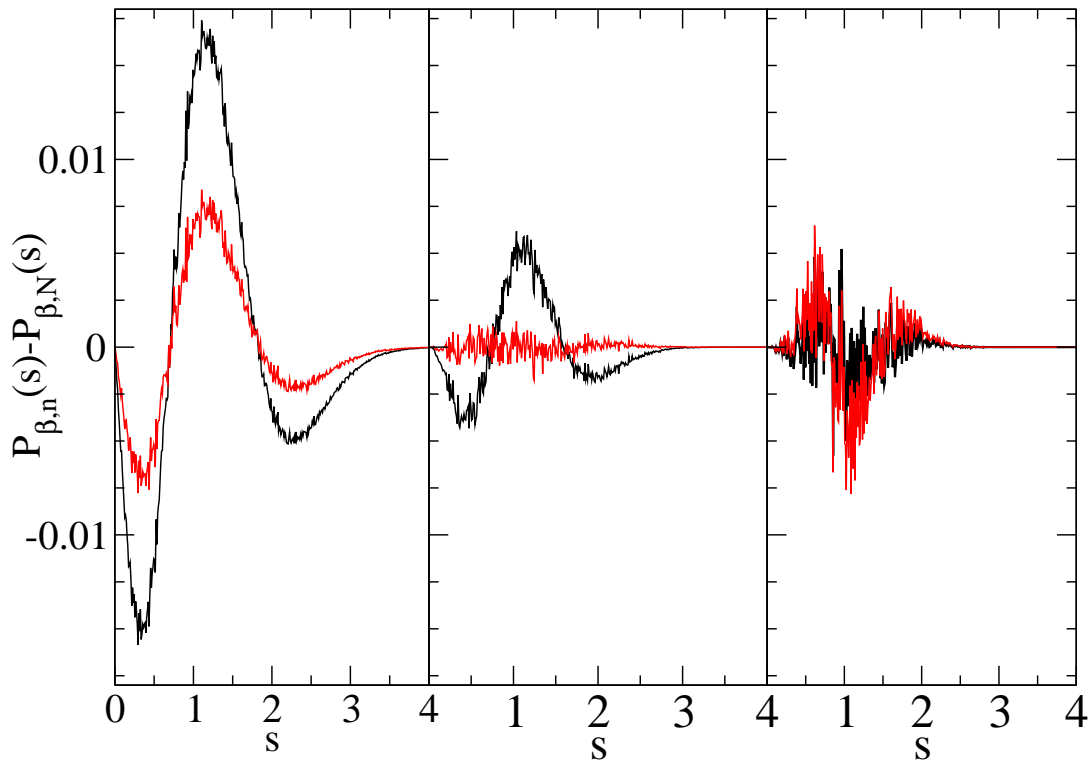}
\caption{Difference between analytic expressions obtained for small $n\times n$ matrices and the nearest-neighbor spacing distribution obtained numerically for $N=1000$ with $n=2$ (Wigner surmise, black) and $n=3$ (analytic expressions \eqref{p1}--\eqref{p4}, red). From left to right: GOE, GUE and GSE. Numerics is obtained from $160000$ matrix realizations for GOE and GUE, $20000$ for GSE.
\label{delta_GUE}}
\end{figure}

It is natural to generalize these well-known arguments by considering not $2\times 2$ matrices but $3\times 3$ ones. Instead of Eq.~\eqref{2_2} one gets 
\begin{equation}
p_{\beta}(s)=\int \tfrac{1}{2}\Big [\delta\big (s-(\lambda_2-\lambda_1)\big )+ \delta\big (s-(\lambda_3-\lambda_2)\big )\Big ]P_{\beta}^{(V)}(\lambda_1,\lambda_2,\lambda_3)\, d\lambda_1 d\lambda_2 d\lambda_3
\end{equation} 
where it is assumed that $\lambda_1\leq \lambda_2\leq \lambda_3$. 
The two terms in this expression are equal and using $\lambda_2=\lambda_1+s$ and $\lambda_3=\lambda_2+y$ the integral over $\lambda_1$ is straightforward and one obtains that up to an overall constant, for $V(\lambda)=\lambda^2$,
\begin{equation}
p_{\beta}(s)=s^{\beta}\int_0^{\infty}y^{\beta}(s+y)^{\beta}\mathrm{e}^{-(s^2+sy+y^2)}dy.
\end{equation} 
The integral can be calculated for integer $\beta$. In particular,
\begin{eqnarray}
p_1(s)&=&\frac{s}{8} \Big[ -\sqrt{\pi}\, \mathrm{erfc}\left ( \frac{s}{2} \right )\,(s^2-2)\, \mathrm{e}^{-3s^2/4}+2s\, \mathrm{e}^{-s^2}\Big ],
\label{p1}\\
p_2(s)&=&\frac{s^2}{32}\Big [\sqrt{\pi}\, \mathrm{erfc}\left ( \frac{s}{2} \right )\,(s^4-4s^2+12)\, \mathrm{e}^{-3s^2/4}
-2s(s^2-6) \mathrm{e}^{-s^2} \Big ],
\label{p2}\\  
p_4(s)&=&\frac{s^4}{512} \Big [\sqrt{\pi}\, \mathrm{erfc}\left ( \frac{s}{2} \right )\,( s^8-8s^6+72s^4 -480s^2+1680)\, \mathrm{e}^{-3s^2/4} -2s(s^6-10s^4+100s^2-840) \mathrm{e}^{-s^2} \Big ] .
\label{p4}   
\end{eqnarray}
Here $\mathrm{erfc}(x)$ is the complementary error function
\begin{equation}
\mathrm{erfc}(x)=\frac{2}{\sqrt{\pi}}\int_x^{\infty}\mathrm{e}^{-t^2}dt .
\end{equation}
Enforcing the normalization conditions \eqref{norm} leads to the following expression for the $3\times 3$ nearest-neighbor distributions 
\begin{equation}
P_{\beta}(s)=a_{\beta}\ p_{\beta}(b_{\beta}s)
\end{equation} 
where constants $a_{\beta}$ and $b_{\beta}$ for $\beta=1,2,4$  have the following values 
\begin{equation}
a_1=\frac{3^3}{2\, \pi},\; b_1=\frac{3}{2\, \pi^{1/2}};\qquad a_2= \frac{3^6}{2^5\, \pi^{3/2}},\; b_2=\frac{3^{5/2}}{2^3\, \pi^{1/2}};\qquad a_4= \frac{3^{10}}{2^{10}5^2\,  \pi^{3/2}},\; b_4=\frac{3^{11/2}}{2^5 5\, \pi^{1/2}}.
\end{equation}
In Fig.~\ref{delta_GUE} the difference between this analytical $P_{\beta}(s)$ and the numerical large-$N$ nearest-neighbor spacing distribution is presented. For comparison, the difference between the usual Wigner surmise \eqref{anzatz} and the exact values are plotted. In the case of GUE, $\beta=2$, the expression \eqref{p2} considerably reduces the error of the approximation, similarly as for the density of ratios.

\section{Joint probability density of the ratios $r_j$ between consecutive nearest-neighbor spacings}
\label{Jointpro}

\subsection{General results}

We now generalize the above results to arbitrary matrix size by considering the more general problem of calculating the joint probability density of ratios for an arbitrary confining potential $V$ and arbitrary $\beta$. Let $\cp_\beta^{(V)} (r_1,\ldots, r_{N-2})$ be the joint probability density of the ratios $r_j=(\lambda_{j+2}-\lambda_{j+1})/(\lambda_{j+1}-\lambda_j)$ between consecutive nearest-neighbor spacings in an $N\times N$ $\beta$-ensemble of random matrices, characterized by the joint probability density of eigenvalues $P_\beta^{(V)} (\lambda_1,\ldots,\lambda_N)$ given by \eqref{P_beta} with confining potential $V(x)$. The eigenvalues have support on an interval $(a,b)$. In the $\beta$-Hermite (Gaussian) case $(V\equiv \mathrm{H})$, we have $a\to -\infty$ and $b\to\infty$, while for $\beta$-Laguerre $(V\equiv \mathrm{L})$ $a\to 0$ and $b\to\infty$.

\subsubsection{Joint density of ratios}

Let us now calculate the joint density of ratios defined by
\begin{equation}
\cp_\beta^{(V)}  (r_1,\ldots, r_{N-2})=\int_{[a,b]^N}d\lambda_1\cdots d\lambda_N P_\beta^{(V)} (\lambda_1,\ldots,\lambda_N)\prod_{j=1}^{N-2}\delta\left(r_j-\frac{\lambda_{j+2}-\lambda_{j+1}}{\lambda_{j+1}-\lambda_j}\right).
\end{equation}
Consider a configuration of $N$ ordered eigenvalues $\lambda_1<\lambda_2<\ldots < \lambda_N$ drawn from the joint density $P_\beta^{(V)}  (\lambda_1,\ldots,\lambda_N)$. By definition, the ratios between consecutive spacings verify 
\begin{align}
\nonumber\lambda_3 &= \lambda_2 + r_1 (\lambda_2 -\lambda_1) \\
 \nonumber\lambda_4 &= r_2 (\lambda_3 -\lambda_2)+\lambda_3  = \lambda_2 + (r_1+r_1 r_2) (\lambda_2-\lambda_1)\\
\lambda_5 &= r_3 (\lambda_4 -\lambda_3)+\lambda_4  = \lambda_2 + (r_1+r_1 r_2+r_1 r_2 r_3) (\lambda_2-\lambda_1)\label{replacem} \\ 
 \nonumber    \vdots &= \vdots
\end{align} 
We introduce the new variables $f_j$ defined by
\begin{equation}
f_j = \sum_{\ell=1}^{j}\prod_{k=1}^\ell r_k,
\label{fj}
\end{equation}
so that for $j\geq 3$ one has 
\begin{equation}
\lambda_j = \lambda_2 + f_{j-2} (\lambda_2-\lambda_1)\,.\label{replacement}
\end{equation}
The $f_j$ auxiliary random variables are sufficiently important in the following to deserve to be given a name. Since from \eqref{replacement}
\begin{equation}
f_j=\frac{\lambda_{j+2}-\lambda_2}{\lambda_2-\lambda_1}\,,\label{RDS}
\end{equation}
we call them \emph{relative disjoint spacings} (RDS). 
 
In order to obtain the joint density of the ratios $r_1,\dots, r_{N-2}$ it is sufficient to replace $\lambda_3,\ldots,\lambda_N$ in $P_\beta^{(V)}  (\lambda_1,\ldots,\lambda_N)$ with the above expression \eqref{replacem}, append a Jacobian factor $||\frac{\partial \lambda_j}{\partial r_k}||$, and integrate over $\lambda_1$ and $\lambda_2$ (the only remaining variables). From \eqref{replacem}, it follows that the $(N-2)\times (N-2)$ Jacobian matrix is lower triangular:
\begin{equation}
\begin{pmatrix}
\frac{\partial \lambda_3}{\partial r_1} & \cdots & \frac{\partial \lambda_3}{\partial r_{N-2}} \\
\vdots & \ddots & \vdots\\
\frac{\partial \lambda_N}{\partial r_1} & \cdots & \frac{\partial \lambda_N}{\partial r_{N-2}} 
\end{pmatrix} =
\begin{pmatrix}
      \lambda_2-\lambda_1 & \multicolumn{2}{c}{\text{\kern0.5em\smash{\raisebox{-1ex}{\Large 0}}}} \\
      (\lambda_2 -\lambda_1)(1+r_2) & r_1 (\lambda_2-\lambda_1) &  \\
     (\lambda_2 -\lambda_1)(1+r_2+r_2 r_3)  &      (\lambda_2 -\lambda_1)(r_1+r_1 r_3)  & r_1 r_2 (\lambda_2 -\lambda_1)\\
     \vdots & \vdots & \ddots
    \end{pmatrix},
\end{equation}
and its determinant (the product of the diagonal elements) is precisely the Jacobian factor
\begin{equation}
\Big|\Big|\frac{\partial \lambda_j}{\partial r_k}\Big|\Big|=\left(\prod_{j=1}^{N-2} r_j^{N-2-j}\right) (\lambda_2-\lambda_1)^{N-2}\,.\label{detjac}
\end{equation}
This yields directly the joint density of ratios

\begin{equation}
\cp_\beta^{(V)}  (r_1,\ldots, r_{N-2})=\left(\prod_{j=1}^{N-2} r_j^{N-2-j}\right)\int_a^b d\lambda_1\int_{\lambda_1}^b d\lambda_2 P_\beta^{(V)}  \left(\lambda_1,\lambda_2,
\{\lambda_j(\mathbf{r})\}\right)(\lambda_2-\lambda_1)^{N-2},
\label{general}
\end{equation}
with $\{\lambda_j(\mathbf{r})\}$ given by the replacement rules \eqref{replacem}. Note that the prefactor $\prod_{j=1}^{N-2} r_j^{N-2-j}$ can be itself conveniently rewritten in terms of the $f_j$, making it clear that the RDS variables, rather than $r_j$ themselves, are the most natural ones in this context. More precisely, one has the identity 
 \begin{equation}
 \prod_{j=1}^{N-2} r_j^{N-2-j}=K(\mathbf{f})\equiv\prod_{j=0}^{N-4}(f_{j+1}-f_j),
 \label{identity}
\end{equation}
where we have introduced $f_0=0$.

The joint density of the RDS is defined as
  \begin{equation}
\label{linkprpf}
  \hat{\cp}_\beta^{(V)} (f_1,\ldots,f_{N-2})  df_1\cdots d f_{N-2}=\cp_\beta^{(V)} (r_1,\ldots,r_{N-2}) d r_1\cdots d r_{N-2}
 \end{equation}
 and for any potential $V(x)$ satisfies the normalization
 \begin{equation}
 \int_0^\infty d f_1 \int_{f_1}^\infty d f_2 \cdots \int_{f_{N-3}}^\infty d f_{N-2}  \hat{\cp}_\beta^{(V)} (f_1,\ldots,f_{N-2})=1\,.\label{normhermite}
 \end{equation}
 Given that the Jacobian factor between the $r_j$ and $f_k$ is independent of $V(x)$ and precisely equal to
 \begin{equation}
\label{jacobrf}
 \Big|\Big|\frac{\partial r_j}{\partial f_k}\Big|\Big| = [K(\mathbf{f})]^{-1},
 \end{equation}
 the joint density of the $f_j$ (using \eqref{general} and \eqref{linkprpf}) reads
 
 \begin{equation}
\hat{\cp}_\beta^{(V)}  (f_1,\ldots, f_{N-2})=\int_a^b d\lambda_1\int_{\lambda_1}^b d\lambda_2 P_\beta^{(V)}  \left(\lambda_1,\lambda_2,
\{\lambda_j(\mathbf{f})\}\right)(\lambda_2-\lambda_1)^{N-2}.
\label{generalf}
\end{equation}

Note that $\hat{\cp}_\beta^{(V)} (f_1,\ldots,f_{N-2})$ is \emph{not} symmetric under the exchange $f_j\to f_k$ (as it may seem at first glance) due to the ordering constraint $f_1\leq f_2\leq \ldots \leq f_{N-2}$, which is reflected in the normalization condition \eqref{normhermite}. However, it is not difficult to define a totally symmetric joint density\footnote{Obviously the Vandermonde term must be then considered in absolute value $\prod_{j<k}^{N-2}(f_k-f_j)^\beta\to\prod_{j<k}^{N-2} |f_k-f_j|^\beta$.} $\tilde{\cp}_\beta^{(\mathrm{V})} (f_1,\ldots,f_{N-2})=\hat{\cp}_\beta^{(\mathrm{V})} (f_1,\ldots,f_{N-2})/(N-2)!$, satisfying the normalization 
 \begin{equation}
  \int_{[0,\infty]^{N-2}} d f_1\cdots d f_{N-2}  \tilde{\cp}_\beta^{(\mathrm{V})} (f_1,\ldots,f_{N-2})=1\,.
\label{normtilde}
 \end{equation}

\subsubsection{Marginal density of ratios: general formula}
Before dealing with more explicit examples for small $N$, our goal here is to present a general integral relation for the density (one-point marginal) $\rho_{\beta,N}^{(V)}(r)$ defined in the standard way as
 \begin{equation}
 \rho_{\beta,N}^{(V)} (r)=\Big\langle \frac{1}{N-2}\sum_{j=1}^{N-2}\delta(r-r_j)\Big\rangle,\label{densitydef}
 \end{equation}
 where the average is over $\cp_\beta^{(V)} (r_1,\ldots, r_{N-2})$. We shall trade the multiple integration over $\cp_\beta^{(V)}  (r_1,\ldots, r_{N-2})$ for a multiple integration over $ \hat{\cp}_\beta^{(V)}  (f_1,\ldots,f_{N-2})$ using \eqref{linkprpf}. The formula presented below will be valid in general for \emph{any} $\beta$-ensemble, and we will later specialize it to the $\beta$-Hermite and $\beta$-Laguerre cases.
We have
\begin{align}
\nonumber \rho_{\beta,N}^{(V)}(r) &=\frac{1}{N-2}\left[ \int_{r\leq f_2\leq \ldots \leq f_{N-2}<\infty} d f_2\cdots d f_{N-2} \hat{\cp}_\beta^{(V)} (r,f_2,\ldots,f_{N-2})\right. \\
\nonumber &+
\int_{0\leq f_1\leq f_1 (r+1)\leq f_3\leq \ldots \leq f_{N-2}<\infty} d f_1 df_3\cdots d f_{N-2} \hat{\cp}_\beta^{(V)} (f_1,f_1 (r+1),f_3,\ldots,f_{N-2}) f_1\\
\nonumber &+
\int_{0\leq f_1\leq f_2\leq (1+r)f_2-r f_1\leq f_4\leq \ldots \leq f_{N-2}<\infty} d f_1 df_2 d f_4\cdots d f_{N-2} \hat{\cp}_\beta^{(V)} (f_1,f_2, (1+r)f_2-r f_1,\ldots,f_{N-2}) (f_2-f_1)\\
&\left. +\ldots +\int_{0\leq f_1\leq \ldots \leq f_{N-3}<\infty} d f_1 \cdots d f_{N-3} \hat{\cp}_\beta^{(V)} (f_1,\ldots,f_{N-3}, (1+r)f_{N-3}-r f_{N-4}) (f_{N-3}-f_{N-4})\right]\,.\label{densityformula1}
\end{align}
More explicitly, we have for instance
\begin{equation}
\int_{0\leq f_1\leq f_2\leq (1+r)f_2-r f_1\leq f_4\leq \ldots \leq f_{N-2}<\infty} d f_1 df_2 d f_4\cdots d f_{N-2} \equiv \int_0^\infty d f_1 \int_{f_1}^\infty d f_2 \int_{(1+r)f_2-r f_1}^\infty d f_4\int_{f_4}^\infty df_5 \cdots 
\end{equation}
Formula \eqref{densityformula1} can be easily proved by expressing $r_j$ as a function of $f_j,f_{j-1}$ and $f_{j-2}$ in \eqref{densitydef} as
\begin{align*}
r_1 &= f_1\\
r_2 &= \frac{f_2-f_1}{f_1}\\
r_3 &= \frac{f_3-f_2}{f_2-f_1}\\
\vdots &= \vdots
\end{align*}
and then use the delta function to kill one of the integrals at a time. For example, we have
\begin{equation}
\delta(r-r_3)=\delta\left(r-\frac{f_3-f_2}{f_2-f_1}\right)=(f_2-f_1)\delta(f_3-((1+r)f_2-r f_1))
\end{equation}
yielding the third line in \eqref{densityformula1}.

We will show how to use \eqref{densityformula1} explicitly to compute the density $\rho_{\beta,N}^{(V)}(r)$ for small $N=3,4,5$ in the next sections, and in particular recover Eqs.~\eqref{surmise3} and \eqref{pderN4-I} from this more general approach. It is also worth mentioning that the knowledge of the full joint distribution of the RDS allows in principle to compute the extreme value statistics for the ratios, such as the cumulative distribution of the maximal ratio

\begin{equation}
\mathbb{P}_{\beta,N}[r_{\mathrm{max}}<x] =\int_{[0,x]^{N-2}} dr_1\cdots d r_{N-2}  \cp_\beta^{(V)} (r_1,\ldots,r_{N-2})
\end{equation}
and its density,
\begin{equation}
p_{\beta,N}^{(V)}(x)=\frac{d}{dx}\mathbb{P}_{\beta,N}[r_{\mathrm{max}}<x].
\end{equation}

\subsection{Application to $\beta$-Hermite ensembles}
\label{betahermite}
 
\subsubsection{Joint density of ratios}
We now apply the general formula \eqref{general} to the $\beta$-Hermite case of a confining potential $V(x)=x^2/2$. The joint probability density of ordered\footnote{The ordering requirement is very important and amounts to an extra factor $N!$ in the denominator of \eqref{zhermite} with respect to the (customary in literature) unordered case.} eigenvalues for the $\beta$-Hermite ensembles reads
\begin{equation}
P_\beta^{(\mathrm{H})} (\lambda_1,\ldots,\lambda_N)=\frac{1}{Z_{N,\beta}^{(\mathrm{H})} } e^{-\frac{1}{2}\sum_{i=1}^N \lambda_i^2}\prod_{j<k}|\lambda_j-\lambda_k|^\beta\,,\label{joint_hermite_eigenvalues}
\end{equation}
where
\begin{equation}
Z_{N,\beta}^{(\mathrm{H})}=\frac{(2\pi)^{N/2}}{N!} \prod_{j=1}^N\frac{\Gamma\left(1+\frac{\beta}{2}j\right)}{\Gamma\left(1+\frac{\beta}{2}\right)}\,. \label{zhermite}
\end{equation}
Let us rewrite the Vandermonde determinant as follows:
\begin{equation}
\prod_{j<k}(\lambda_j-\lambda_k) =(\lambda_1-\lambda_2)\prod_{j=3}^N (\lambda_1-\lambda_j)(\lambda_2-\lambda_j)\prod_{3\leq j<k}(\lambda_j-\lambda_k).
\end{equation}
Now it is clear that upon the replacement rule \eqref{replacement}, the Vandermonde interaction term $\prod_{j<k}|\lambda_j-\lambda_k|^\beta$ will always come in a factorized form
\begin{equation}
\prod_{j<k}|\lambda_j-\lambda_k|^\beta=(\lambda_2-\lambda_1)^{\frac{\beta N (N-1)}{2}}\prod_{j=1}^{N-2} f_j^\beta (1+f_j)^\beta \prod_{j<k}^{N-2}(f_k-f_j)^\beta\label{factorized}
\end{equation}
where the $f_j$, defined in \eqref{factorized}, are positive and such that $f_k>f_j$ for $k>j$.
  
 We are now ready to perform the $(\lambda_1,\lambda_2)$ integration in \eqref{general} that reads
 \begin{align}
\nonumber & \int_{-\infty}^\infty d\lambda_1\int_{\lambda_1}^\infty d\lambda_2 (\lambda_2-\lambda_1)^{N-2+\frac{\beta}{2}N(N-1)}\exp\left[-\frac{1}{2}\left(\lambda_1^2+\lambda_2^2+\sum_{j=3}^{N} (\lambda_2+(\lambda_2-\lambda_1)f_j)^2\right)\right]\\
&=\frac{\sqrt{\frac{\pi}{N}}2^{-1+\frac{N}{4}(2+\beta(N-1))}\Gamma(q)N^q}{\left[N+N\sum_j f_j^2-1-\left(\sum_j f_j\right)^2+2\sum_j f_j\right]^q}
 \end{align}
 where
 \begin{equation}
 q=\frac{(N-1)(2+\beta N)}{4}\,.
 \end{equation}

Using Eqs.~\eqref{linkprpf}--\eqref{jacobrf} we get the joint distribution for the RDS in the $\beta$-Hermite case:
 \begin{equation}
\hat{\cp}_\beta^{(\mathrm{H})} (f_1,\ldots,f_{N-2})=\frac{1}{Z_{N,\beta}^{(\mathrm{H})}}\prod_{j=1}^{N-2} f_j^\beta (1+f_j)^\beta \prod_{j<k}^{N-2}(f_k-f_j)^\beta \frac{\sqrt{\frac{\pi}{N}}2^{-1+\frac{N}{4}(2+\beta(N-1))}\Gamma(q)N^q}{\left[N+N\sum_j f_j^2-1-\left(\sum_j f_j\right)^2+2\sum_j f_j\right]^q}\label{finalgaussian}
 \end{equation}
 where $Z_{N,\beta}^{(\mathrm{H})}$ is given by \eqref{zhermite}.
 This is one of the main results of this paper. 

In the variables $f_j$, the associate joint probability $ \tilde{\cp}_\beta^{(\mathrm{H})} (f_1,\ldots,f_{N-2})$ given by \eqref{normtilde} resembles the joint probability density of eigenvalues of a (non classical) $\beta$-ensemble with a peculiar `fat-tailed' confining potential. Similar ensembles have been considered previously in the literature \cite{wd1,wd2,wd3,wd4,wd5} , where it was noticed that the use of suitable integral identities allows to compute analytically correlation functions for both finite and large $N$. It would be interesting to see if such analytical tools (or variations thereof) could be of any use to compute in a compact form correlation functions of the $f_j$, or equivalently of the $r_j$. Once the dependence on the $r_j$ is explicitly restored, the resulting joint probability density $\cp_\beta^{(\mathrm{H}) }(r_1,\ldots, r_{N-2})$ is \emph{not} symmetric under the exchange $r_j\to r_k$ and satisfies the general duality relation
 \begin{equation}
 \cp_\beta^{(\mathrm{H}) }\left(\frac{1}{r_{N-2}},\ldots, \frac{1}{r_1}\right)=r_1^2\cdots r_{N-2}^2 \cp_\beta^{(\mathrm{H}) } (r_1,\ldots, r_{N-2}),\label{duality}
 \end{equation}
 the higher-dimensional analogue of \eqref{duality3}.

\subsubsection{Marginal density of ratios for small-size matrices $N=3,4,5$}
\label{calculHn4}
The marginal density of ratios can be obtained from the joint density \eqref{finalgaussian} using \eqref{densityformula1}. For the $\beta$-Hermite ensemble (for any $N$ and $\beta$) it satisfies again the duality relation \eqref{duality3}. Let us now consider small-$N$ cases. 

\emph{The case $N=3$. }
In this case, Eq.~\eqref{finalgaussian} directly gives the distribution of $f_1=r$ (the only variable). Explicitly, it yields
\begin{equation}
\hat{\cp}_\beta^{(\mathrm{H})}(r)=\rho_{\beta,3}^{(\mathrm{H})}(r)=\frac{1}{Z_\beta}\frac{(r+r^2)^{\beta}}{(1+r+r^2)^{1+3\beta/2}},
\end{equation}
while Eq.~\eqref{zhermite} gives
\begin{equation}
Z_\beta =\frac{2\pi\Gamma(1+\beta)}{3^{3(1+\beta)/2}\Gamma(1+\beta/2)^2}\,,
\end{equation}
recovering exactly the surmise \eqref{surmise3}. The asymptotic behavior of the density is given by $\rho_{\beta,N}^{(\mathrm{H})}(r) \sim r^\beta$ for $r\to 0$ and $\rho_{\beta,N}^{(\mathrm{H})}(r) \sim r^{-2-\beta}$ for $r\to \infty$. In the subsequent cases, we will see that this asymptotic behavior is quite robust, as announced in \eqref{asbeh}.

\emph{The case $N=4$. }
In this case, formula \eqref{densityformula1} for the density explicitly reads
\begin{equation}
 \rho_{\beta,N}^{(V)}(r) =\frac{1}{2}\left[\int_r^\infty df_2  \hat{\cp}_\beta^{(\mathrm{H}) }\left(r,f_2\right)+\int_0^\infty df_1  \hat{\cp}_\beta^{(\mathrm{H}) }\left(f_1,f_1(r+1)\right)\right],
\end{equation}
yielding

\begin{align}
\nonumber \rho_{\beta,4}^{(\mathrm{H})}(r) &=\frac{\sqrt{\pi} \Gamma\left(\frac{3}{2}(1+2\beta)\right)2^{3+9\beta} r^\beta (1+r)^\beta}{2 Z_{4,\beta}^{(\mathrm{H})}}\left[\int_r^\infty d f_2 \frac{f_2^\beta (1+f_2)^\beta (f_2-r)^\beta}{\left[
3+4(r^2+f_2^2)-(r+f_2)^2+2(r+f_2)\right]^{3(1+2\beta)/2}}+\right.\\
&\left.+\int_0^\infty d f_1 \frac{f_1^{3\beta+1} (1+f_1)^\beta (1+f_1(1+r))^\beta}{\left[
3+4 f_1^2(1+(1+r)^2)-f_1^2(2+r)^2+2 f_1(2+r)\right]^{3(1+2\beta)/2}}\right]\,.\label{densityN4}
\end{align}
The integrals can then be easily computed analytically for $\beta=1,2,4$. For $\beta=2$ the expression obtained from \eqref{densityN4} coincides with the one given by \eqref{pderN4-I}--\eqref{pderN4-II}. For $\beta=1$ and $\beta=4$, explicit formulas can be obtained from \eqref{densityN4} and yield expressions of a form similar to \eqref{pderN4-I}--\eqref{pderN4-II}. The asymptotic behavior is again given by \eqref{asbeh}.

\emph{The case $N=5$. }
In the case $N=5$, again, explicit formulas are easy to derive in the same way, but are too long to be reported here. We checked that the asymptotic behavior is again given by \eqref{asbeh}.

\begin{figure}[!t]
\begin{center}
\centering \includegraphics*[width=0.8\linewidth]{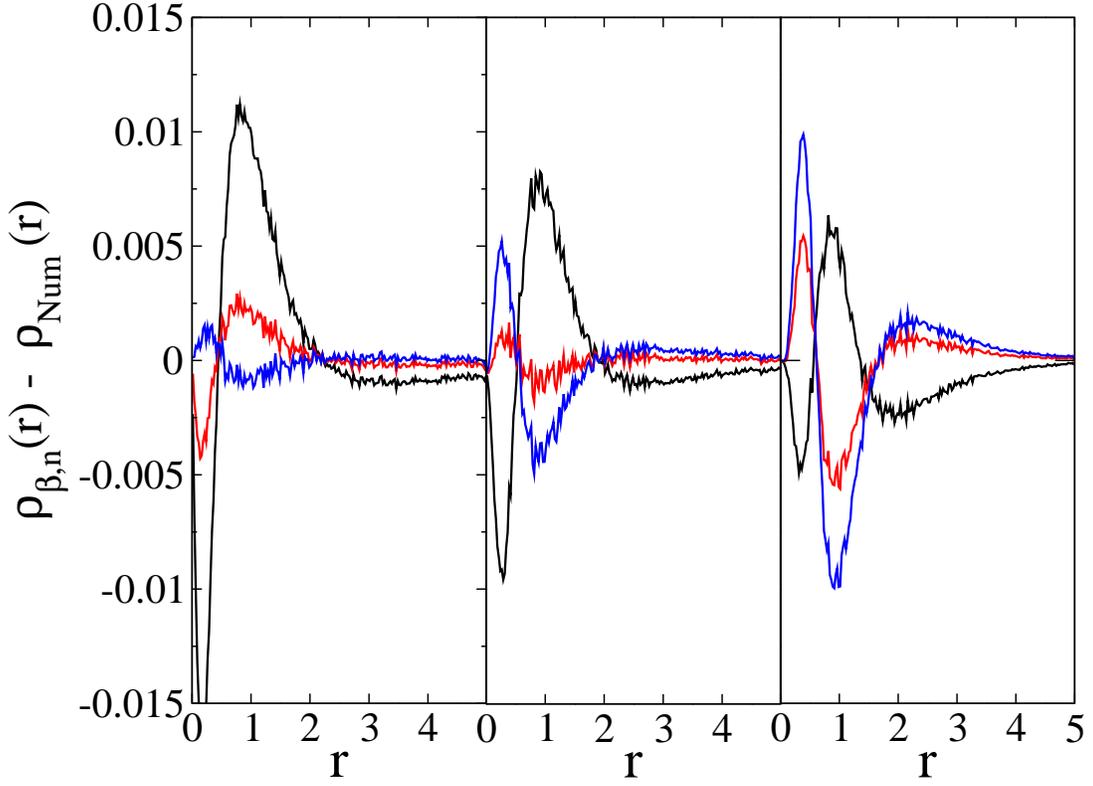}
\caption{Difference between analytic expressions obtained for small $n \times n$ matrices and the distribution of consecutive ratio obtained
numerically for $N = 1000$, with $n = 3$ (Eq. \eqref{surmise3}, black), $n = 4$ (analytic expressions \eqref{densityN4}, red), and $n=5$ (blue) (from bottom to top at $r=3$). From left to right: GOE, GUE and GSE. Numerics is obtained from the full spectrum of 160000 matrix realizations of tridiagonal matrices (see \cite{dumitriu}).  
\label{densityN4fig}}
\end{center}
\end{figure}

In Fig.~\ref{densityN4fig} we show the analytical result obtained for $N=3,4,5$ together with numerical simulations for large-$N$ matrices. These results clearly show that the discrepancy between the small-$N$ and the large-$N$ case for $\beta$-Hermite ensembles is quite small, and thus the small-$N$ formulae can serve as a very good approximation to the asymptotic expressions.  

 \subsection{Application to $\beta$-Laguerre ensembles}\label{betalaguerre}
 
 The $\beta$-Laguerre ensemble is characterized by the joint density of non-negative ordered eigenvalues
 \begin{equation}
 P_\beta^{(\mathrm{L})}(\lambda_1,\ldots,\lambda_N)=\frac{1}{Z_{N,\alpha,\beta}^{(\mathrm{L})}}\prod_{j<k}|\lambda_j-\lambda_k|^\beta \prod_{j=1}^N e^{-\frac{1}{2}\lambda_j}\lambda_j^{\alpha-1}\,,\label{eqWishart}
 \end{equation}
 where
 \begin{equation}
 Z_{N,\alpha,\beta}^{(\mathrm{L})}=\frac{2^{\alpha N +\frac{\beta}{2}N(N-1)}}{N!
}\prod_{j=1}^N \frac{\Gamma\left(\alpha+(j-1)\frac{\beta}{2}\right)\Gamma\left(1+j\frac{\beta}{2}\right)}{\Gamma\left(1+\frac{\beta}{2}\right)}
 \end{equation}
 is the normalization constant.
 
 For $\beta=1,2,4$ the explicit realization of \eqref{eqWishart} is possible in terms of so-called Wishart matrices $\mathbf{W}=\mathbf{X}^\dagger\mathbf{X}$, where $\mathbf{X}$ is a rectangular $M\times N$ ($M\geq N$)
 matrix whose entries are filled with Gaussian random variables (real, complex or quaternions) with mean $0$ and variance $1$. In this case, we have $\alpha=(\beta/2)(M-N+1)$. For illustrative purposes, we confine ourselves to the case $\alpha=1$.

 \subsubsection{Joint density of ratios}
 The double integral \eqref{general} can be performed with $a\to 0$ and $b\to\infty$ and the final result for the associated joint density $ \hat{\cp}_\beta^{(\mathrm{L})} (f_1,\ldots,f_{N-2})$ reads
 \begin{equation}
\hat{\cp}_\beta^{(\mathrm{L})} (f_1,\ldots,f_{N-2})=\frac{1}{Z_{N,1,\beta}^{(\mathrm{L})}}\frac{\Gamma(s-1)2^s}{N}\prod_{j=1}^{N-2} f_j^\beta (1+f_j)^\beta \prod_{j<k}(f_k-f_j)^\beta \left(N-1+\sum_j f_j\right)^{1-s}\,,\label{finalwishart}
 \end{equation}
where
\begin{equation}
s=N+\frac{\beta N (N-1)}{2}\,.
\end{equation}
 
 Once again, we have the normalization condition \eqref{normhermite}. Note that due to the lack of translational invariance (there is a hard wall at $0$) the duality relation \eqref{duality} no longer holds. One could again symmetrize \eqref{finalwishart} defining a new joint density  $\tilde{\cp}_\beta^{(\mathrm{L})} (f_1,\ldots,f_{N-2})=\hat{\cp}_\beta^{(\mathrm{L})} (f_1,\ldots,f_{N-2})/(N-2)!$, satisfying the normalization \eqref{normtilde}. In the variables $f_j$ the new joint density $\tilde{\cp}_\beta^{(\mathrm{L})} (f_1,\ldots,f_{N-2})$ looks very similar to the joint density of eigenvalues of power-law deformations of Wishart matrices \cite{lag1,lag2,lag3}. In particular, given that
\begin{equation}
 \left(N-1+\sum_j f_j\right)^{1-s}\to \exp\left(-\frac{\beta}{2}N \sum_j f_j\right),\qquad\mbox{for }N\to\infty
\end{equation}
 we expect a (rather slow) convergence of the one-point density of the $f_j$ to the Mar\v cenko-Pastur distribution, with strong finite-size corrections that may be perhaps studied using deformed Laguerre polynomials \cite{forrdeformed}. 

In the following subsections, we will investigate in more details the one-point density for the ratios, and we will show how to proceed with a systematic evaluation of averages over $\hat{\cp}_\beta^{(\mathrm{L})} (f_1,\ldots,f_{N-2})$ for the representative case of $\langle r\rangle$ and $N=5,\beta=1$.

\subsubsection{Marginal density of ratios for small-size matrices $N=3,4,5$}

\emph{The case $N=3$. }
In this case, we have $s=3+3\beta$ and Eq. \eqref{finalwishart} for $f_1=r$ directly gives
\begin{equation}
\hat{\cp}_\beta^{(\mathrm{L})}(r)=\rho_{\beta,3}^{(\mathrm{L})}(r)=\frac{4^{1+\beta}\Gamma\left(\frac{3(1+\beta)}{2}\right)}{\Gamma\left(\frac{1+\beta}{2}\right) \Gamma(1+\beta)}\frac{(r+r^2)^\beta}{(2+r)^{2+3\beta}}\label{rhoLN3eq}
\end{equation}
with asymptotic behaviors again of the form \eqref{asbeh}.

\emph{The case $N=4$. }
Formula \eqref{densityformula1} for the density explicitly reads
\begin{equation}
 \rho_{\beta,4}^{(\mathrm{L})}(r) =\frac{\Gamma(3+6\beta)2^{4+6\beta}}{8}\frac{r^\beta (1+r)^\beta}{Z_{4,1,\beta}^{(\mathrm{L})}}\left[\int_r^\infty d f_2 \frac{f_2^\beta (1+f_2)^\beta (f_2-r)^\beta}{\left(
3+r+f_2\right)^{3+6\beta}}+
\int_0^\infty d f_1 \frac{f_1^{3\beta+1} (1+f_1)^\beta (1+f_1(1+r))^\beta}{\left(
3+f_1(2+r)\right)^{3+6\beta}}\right]\,.\label{densN4L}
\end{equation}
The integrals are difficult to compute for general $\beta$. However it is possible to give explicit expressions for $\beta=1,2,4$. For example for $\beta=1$ we get
\begin{equation}
\rho_{1,4}^{(\mathrm{L})}(r)= \frac{32}{9}r(1+r)\left[\frac{23+23 r+2 r^2}{(2+r)^7}+\frac{9(42+91 r+47 r^2)}{(3+2 r)^7}\right],
\end{equation}
confirming the universal asymptotic behavior \eqref{asbeh} for $r\to 0$ and $r\to\infty$.

 \emph{The case $N=5$. }
Again similar expressions can be derived in that case. Rather than giving the explicit equations, it is more instructive to see now how the knowledge of the full joint probability density of the $f_j$ allows to perform nontrivial calculations, such as the average $\langle r\rangle$ for (in principle) any fixed $N$ and $\beta$ integer. For simplicity, we stick to the $\beta$-Laguerre case with $N=5$ and $\beta=1$. By definition
 \begin{equation}
 \langle r\rangle =\frac{1}{3}\langle(r_1+r_2+r_3)\rangle=\frac{1}{3}\Big\langle\left(f_1+\frac{f_2-f_1}{f_1}+\frac{f_3-f_2}{f_2-f_1}\right)\Big\rangle
 \end{equation} 
 where the average is taken over the joint density \eqref{finalwishart}. What we have to compute is then
   \begin{equation}
 \langle r\rangle =\frac{1}{3}\int_0^\infty d f_1 \int_{f_1}^\infty d f_2\int_{f_2}^\infty d f_3  \hat{\cp}^{(\mathrm{L})}_1 (f_1,f_2,f_3) \left(f_1+\frac{f_2-f_1}{f_1}+\frac{f_3-f_2}{f_2-f_1}\right)
 \end{equation}
 and similar expressions for higher $N$ and different $\beta$. The integrand involves the product
 \begin{equation}
 f_1 f_2 f_3 (1+f_1)(1+f_2)(1+f_3)(f_2-f_1)(f_3-f_1)(f_3-f_2)\left(f_1+\frac{f_2-f_1}{f_1}+\frac{f_3-f_2}{f_2-f_1}\right).\label{productf}
 \end{equation}
  One notices that the denominators simplify (this feature remains for any higher $N$ and for any integer $\beta$) and eventually yields after expanding \eqref{productf} a finite sum of terms of the form $f_1^{m_1}f_2^{m_2}f_3^{m_3}$ with $m_1,m_2,m_3$ integers. For general $N$ one has of course $f_1^{m_1}\cdots f_{N-2}^{m_{N-2}}$. All one has to do (see Eq. \eqref{finalwishart}) is then to compute integrals of the form
 \begin{equation}
 \mathcal{I}(m_1,m_2,\ldots,m_{N-2})=\int_0^\infty d f_1\int_{f_1}^\infty d f_2\cdots \int_{f_{N-3}}^\infty d f_{N-2} \frac{f_1^{m_1}\cdots f_{N-2}^{m_{N-2}}}{ \left(N-1+\sum_j f_j\right)^{s-1}}\label{intproduct}
 \end{equation}
and then sum them up with appropriate coefficients arising from the expansion \eqref{productf}. The case $N=5$ will be carried out in detail and should convince the reader that the extension to the general $N$ case does not present further conceptual difficulties.
 
Let us start by representing the denominator in the integrand of \eqref{intproduct} for $N=5$ as
\begin{equation}
\frac{1}{\left(N-1+\sum_j f_j\right)^{s-1}}=\frac{1}{(N-1)^{s-1}\Gamma(s-1)}\int_0^\infty d\xi \xi^{s-2}e^{-\xi}e^{-\frac{\xi}{N-1}\sum_j f_j}
\end{equation}
 and let us perform the $f_3$ (rightmost) integration after shifting $f_3=f_2+\tau_3$:
 \begin{equation}
 \int_{f_2}^\infty d f_3 e^{-\frac{\xi}{N-1}f_3} f_3^{m_3}=\sum_{k_3=0}^{m_3}{m_3 \choose k_3} f_2^{k_3} e^{-\frac{\xi}{N-1}f_2}\underbrace{\int_0^\infty d\tau_3 e^{-\frac{\xi}{N-1}\tau_3}\tau_3^{m_3-k_3}}_{\Gamma(1+m_3-k_3)\left(\frac{\xi}{N-1}\right)^{-1-m_3+k_3}}
 \end{equation}
 and the dependence on $f_2$ and $f_3$ has completely decoupled. Next we can perform the $f_2$ integration, after shifting $f_2 = f_1+\tau_2$:
 \begin{equation}
  \int_{f_1}^\infty d f_2 e^{-\frac{2\xi}{N-1}f_2} f_2^{m_2+k_3}  =\sum_{k_2=0}^{m_2+k_3}{m_2+k_3 \choose k_2} f_1^{k_2} e^{-\frac{2\xi}{N-1}f_1}\underbrace{\int_0^\infty d\tau_2 e^{-\frac{2\xi}{N-1}\tau_2}\tau_2^{m_2+k_3-k_2}}_{\Gamma(1+m_2+k_3-k_2)\left(\frac{2\xi}{N-1}\right)^{-1-m_2-k_3+k_2}}
 \end{equation}
  and eventually
   \begin{equation}
  \int_{0}^\infty d f_1 e^{-\frac{3\xi}{N-1}f_1} f_1^{m_1+k_2}  =\Gamma(1+m_1+k_2)\left(\frac{3\xi}{N-1}\right)^{-1-m_1-k_2}
 \end{equation}
  so that 
  \begin{align}
\nonumber \mathcal{I}(m_1,m_2,m_3)&=\frac{1}{4^{14}\Gamma(14)} \sum_{k_3=0}^{m_3}{m_3 \choose k_3} \sum_{k_2=0}^{m_2+k_3}{m_2+k_3 \choose k_2} \Gamma(1+m_3-k_3)\Gamma(1+m_2+k_3-k_2)\Gamma(1+m_1+k_2)\times\\
\nonumber &\times 2^{-1-m_2-k_3+k_2} 3^{-1-m_1-k_2}4^{3+m_1+m_2+m_3}\int_0^\infty d\xi e^{-\xi}\xi^{10-(m_1+m_2+m_3)}=\\
\nonumber &=\frac{1}{4^{14}\Gamma(14)} \sum_{k_3=0}^{m_3}{m_3 \choose k_3} \sum_{k_2=0}^{m_2+k_3}{m_2+k_3 \choose k_2} \Gamma(1+m_3-k_3)\Gamma(1+m_2+k_3-k_2)\Gamma(1+m_1+k_2)\times\\
&\times 2^{-1-m_2-k_3+k_2} 3^{-1-m_1-k_2}4^{3+m_1+m_2+m_3}\Gamma(11-(m_1+m_2+m_3)).\label{generalintegration}
 \end{align}
 
 We stress that a general formula for any desired $N$ is not difficult to obtain given the iterative decoupling of integration variables as shown above for the case $N=5$. The resulting formula will involve $N-3$ nested sums. For $N=5$ and $\beta=1$ using the reduction formula above we obtain the exact analytic expression
\begin{equation}
\langle r\rangle =\frac{175271}{52 488}\approx 3.33926...
\end{equation}
We also believe that a similar expression might be worked out for the Gaussian case also, even though the presence of the square in the exponents may require some more work.

\section{Distribution of the $k$th overlapping ratios}
\label{sec.overlap}
 
\begin{figure}[!b]
\centering \includegraphics*[width=0.6\linewidth]{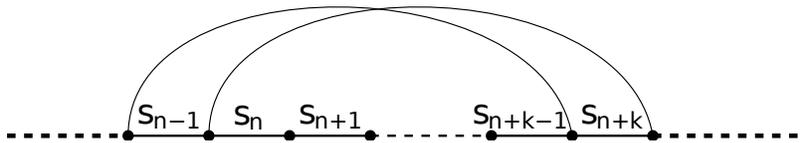}
\caption{$k$th overlapping ratio. The number $k$ represents the number of shared spacings.\label{overlaps}}
\end{figure}

The previous section was devoted to the calculation of joint distributions for non-overlapping ratios of level spacings. We now turn to a different kind of generalization of the ratios of successive spacings, namely the $k$th overlapping ratios introduced in \cite{atas}, which we  define as
\begin{equation}
r_{n}^{(k)}=\frac{s_{n}+s_{n+1}+\cdots s_{n+k}}{s_{n-1}+s_{n}+\cdots +s_{n+k-1}}=\frac{\lambda_{n+k+1}-\lambda_n}{\lambda_{n+k}-\lambda_{n-1}}, \label{k_overlapping_def}
\end{equation}
with $k$ the number of shared spacings (see Fig.~\ref{overlaps}). It can be used in the same way as the ratio $r_{n}=s_{n+1}/s_{n}$ studied in the previous sections for comparison with numerical data. In the large $k$ limit, one expects that the distribution of the $r_{n}^{(k)}$ will be peaked around $1$. 

 \subsection{Poisson case} 
We first compute the distribution $P_{k}(r)$ of the quantity $r_n^{(k)}$ for Poisson random variables $\lbrace\lambda_{i}\rbrace$. All the calculation can be done exactly for any value of $k$. The distribution $P_{k}(r)$ is given by:
\begin{equation}
P_{k}(r)=\int_0^\infty \prod_{j=n-1}^{n+k} ds_{j} \mathrm{e}^{ -s_{n-1}-s_{n}-\cdots -s_{n+k}}\delta \left( r-\frac{s_{n}+s_{n+1}+\cdots +s_{n+k}}{s_{n-1}+s_{n}+\cdots +s_{n+k-1}}\right). \label{overlap_k_integral}
\end{equation}
Let us denote $x=s_{n}+s_{n+1}+\cdots +s_{n+k-1}$ the size of the interval shared by the numerator and the denominator. Equation (\ref{overlap_k_integral}) then becomes:
\begin{equation}
P_{k}(r)=\int  ds_{n-1} ds_{n+k}dx \hspace*{0.2cm} p_{k}(x)\mathrm{e}^{ -s_{n-1}-s_{n+k}}\delta \left( r-\frac{x+s_{n+k}}{s_{n-1}+x}\right), \label{overlap_k_integral2}
\end{equation}
where
\begin{equation}
p_{k}(x)=\frac{x^{k-1}}{(k-1)!}\mathrm{e}^{-x}
\end{equation}
can be obtained either by direct integration in \eqref{overlap_k_integral} or by noticing that it
represents the probability of observing an interval of size $x$ containing $k$ eigenvalues. 
Inserting this expression in (\ref{overlap_k_integral2}) and integrating over $x$ leads to:
\begin{equation}
P_{k}(r)=\frac{1}{(k-1)! (r-1)^{k+1}} \int ds_{n-1}ds_{n+k}\left(s_{n+k}-rs_{n-1} \right)^{k-1}\mathrm{exp}\left(-\frac{1}{r-1}(s_{n+k}r-s_{n-1})\right)(s_{n+k}-s_{n-1}),
\end{equation}
with the condition 
\begin{equation}
\frac{s_{n+k}-rs_{n-1}}{r-1}\geq 0.
\end{equation}
These integrals can be easily performed and one finds:
\begin{equation}
P_{k}(r)=\left\{
\begin{array}{c}
  \dfrac{r^{k}(k+1+kr)}{(1+r)^2}
\hspace{0.2cm} \mathrm{if} \hspace{0.2cm} r<1,  \vspace*{0.2cm}\\
  \dfrac{k+r(k+1)}{r^{k+1}(1+r)^2} \hspace{0.2cm} \mathrm{if} \hspace{0.2cm} r>1.
\end{array}
\right. \label{k-recouvrement}
\end{equation}
For $r\rightarrow 0$ one has the asymptotic behavior $P_{k}(r) \sim r^{k+1}$, and for $r\rightarrow \infty$ the asymptotic reads $P_{k}(r) \sim r^{-k-2}$. 

\begin{figure}[!t]
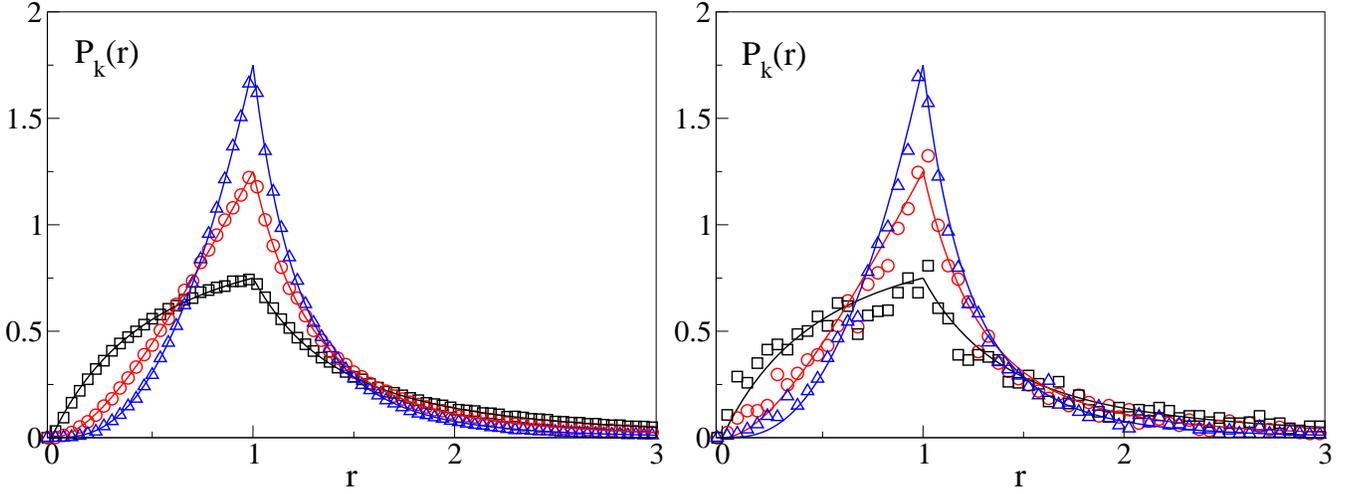

\centering \includegraphics*[width=0.49\linewidth]{fig4a}
\centering \includegraphics*[width=0.49\linewidth]{fig4b}
\caption{Left: histogram of the $k$th overlapping ratio distribution for 355443 lowest energy levels of a rectangular billiard of size $2^{1/4}\times 5^{1/4}$ for $k=1$ (black squares), $k=2$ (red circles) and $k=3$ (blue triangles). Full lines correspond to formula (\ref{k-recouvrement}). Right: $k=1$ (black squares), $k=2$ (red circles) and $k=3$ (blue triangles) overlapping ratio distribution for the near-integrable quantum Ising model  for $\lambda=1$, $\alpha=0.01$ and $L=18$ spins in the sector of eigenvalue $\omega_{4}$ of the translation operator $\hat{T}$.\label{rectbilliard} }
\end{figure}

It was conjectured by Berry and Tabor in 1977 \citep{berry} that the distribution of eigenvalues for quantum systems with integrable classical  counterpart is that of a Poisson process. We first consider the energy levels of a free particle in an incommensurate rectangular billiard of size $a \times b$ with periodic boundary conditions:
\begin{equation}
\lambda_{l,m}=\left( \frac{2\pi l}{a}\right)^{2}+\left( \frac{2\pi m}{b}\right)^{2},  \hspace*{0.2cm} l,m=0,1,\dots \label{Energy_billiard}
\end{equation}
The overlapping ratio distribution calculated from the levels in \eqref{Energy_billiard} is presented in Fig.~\ref{rectbilliard}, showing perfect agreement with (\ref{k-recouvrement}). 

The second example comes from a quantum many-body lattice problem. The one-dimensional quantum Ising model with $L$ spins in transverse and longitudinal fields with periodic boundary conditions is defined by the following Hamiltonian:
\begin{equation}
\label{IsingHamiltonian}
\hat{H}=-\sum_{n=1}^{L}\left( \hat{\sigma}_{n}^{x}\hat{\sigma}_{n+1}^{x}+\lambda \hat{\sigma}_{n}^{z}+\alpha \hat{\sigma}_{n}^{x}\right), \hspace{0.2cm} \hat{\sigma}_{L+1}^{x}=\hat{\sigma}_{1}^{x}, 
\end{equation} 
with $\hat{\sigma}_{n}^{x,z}$ the usual Pauli matrices. For $\alpha=0$, this model reduces to the integrable quantum Ising chain, which is the canonical model of a quantum phase transition \cite{pfeuty,sachdev}.
The Hamiltonian commutes with the translation operator $\hat{T}$ which translates the state by one lattice spacing. Obviously, $\hat{T}^{L}=1$, thus eigenvalues of $\hat{T}$ are given by the $L$ roots of unity $\omega
_{j}=\exp( 2\mathrm{i}\pi j/L), \hspace*{0.2cm}j=0,1,\dots, L-1$. $\hat{H}$ takes a block diagonal form in the basis of eigenstates of $\hat{T}$ and one has to consider energies of $\hat{H}$ in the different sectors of symmetry labeled by the quantum number $\omega_{j}$. The result for one sector is presented in Fig.~\ref{rectbilliard}. Numerical results for $P_{1}(r)$ and $\alpha \rightarrow 0$ agree very well with the Poisson prediction.

\subsection{k=1 overlapping ratio distribution for the $\beta$-Hermite ensemble}
We now compute the $k=1$ overlapping ratio distribution for the $\beta$-Hermite ensembles using an argument similar to the one discussed in Section \ref{small34}. We expect that a good surmise should be given by the distribution for the smallest matrix size which is $N=4$. 
The distribution of the quantity $r=(\lambda_{4}-\lambda_{2})/(\lambda_{3}-\lambda_{1})$ is given by
\begin{equation}
P_{\beta}(r)=\int \prod_{i=1}^{4}d\lambda_{i} P_{\beta}^{(\mathrm{H})}(\lambda_{1},\lambda_{2},\lambda_{3},\lambda_{4})\,\delta\left(r-\frac{\lambda_{4}-\lambda_{2}}{\lambda_{3}-\lambda_{1}}\right), \label{beta_hermite_overlap_def}
\end{equation}
with $ P_{\beta}^{(\mathrm{H})}(\lambda_{1},\lambda_{2},\lambda_{3},\lambda_{4})$ the joint probability distribution \eqref{P_beta} of four eigenvalues in the $\beta-$Hermite ensembles.
Let us introduce the spacing variables
\begin{equation}
\lambda_{1}=\lambda, \hspace{0.2cm} \lambda_{i}=\lambda +\sum_{j=1}^{i-1}s_{j}, \hspace*{0.2cm} i=2,3,4,
\end{equation}
so that 
\begin{equation}
P_{\beta}(r)= \int_{-\infty}^{\infty} d\lambda\int_{0}^{\infty}ds_{1}ds_{2}ds_{3} P_{\beta}^{(\mathrm{H})}(\lambda,\lambda+s_{1},\lambda+s_{1}+s_{2},\lambda+s_{1}+s_{2}+s_{3})\, \delta\left(r-\frac{s_{2}+s_{3}}{s_{1}+s_{2}}\right),
\end{equation}
with $P_{\beta}(\lambda_{1},\dots,\lambda_{4})$ given by \eqref{joint_hermite_eigenvalues}. The integral over $\lambda$ reads
\begin{equation}
\int_{-\infty}^{\infty}d\lambda\ \mathrm{exp}\left( -2\lambda^{2}+\lambda( 3s_{1}+2 s_{2}+ s_{3})\right)=\sqrt{\frac{\pi}{2}}\mathrm{exp}\left(\frac{1}{8}( 3s_{1}+2 s_{2}+ s_{3})^{2}\right).
\end{equation}
Integrating over $s_{2}$ we obtain
\begin{align}
\notag P_{\beta}(r)=\frac{C_{\beta}|r|^{\beta}}{|r-1|^{4\beta +2}}\int_{\frac{s_{3}-rs_{1}}{r-1}\geq 0}ds_{1}ds_{3}\ &s_{1}^{\beta}s_{3}^{\beta}|(s_{3}-rs_{1})(rs_{3}-s_{1})|^{\beta}|s_{3}-s_{1}|^{2\beta+1}\\
& \times \exp\left( -\frac{1}{8(r-1)^{2}}(h(r)(s_{1}^{2}+s_{3}^{2})-2(1+r)^{2}s_{1}s_{3})\right) ,
\end{align}
with $h(r)=3-2r+3r^{2}$ and $C_{\beta}$ a normalization constant. The integrand is invariant under the permutation $s_{1} \leftrightarrow s_{3}$. Introducing a dimensionless variable $z=s_{1}/s_{3}$ one finally obtains
\begin{equation}
P_{\beta}(r)=
\begin{cases}
c_{\beta}|r|^{\beta}(1-r)^{2\beta+1}\displaystyle{\int_{0}^{r}}\dfrac{z^{\beta}(1-rz)^{\beta}(r-z)^{\beta}(1-z)^{2\beta+1}}{\left( h(r)(1+z^{2})-2(1+r)^{2}z\right)^{3/2+3\beta}} dz & \mathrm{for} \hspace*{0.2cm} r<1 \\
&\\
c_{\beta}|r|^{\beta}(r-1)^{2\beta+1}\displaystyle{\int_{r}^{\infty}}\dfrac{z^{\beta}(rz-1)^{\beta}(z-r)^{\beta}(z-1)^{2\beta+1}}{\left( h(r)(1+z^{2})-2(1+r)^{2}z\right)^{3/2+3\beta}}dz & \mathrm{for} \hspace*{0.2cm} r>1, \label{result_final_hermite_rgrand}
\end{cases}
\end{equation}
where $c_{\beta}$ is a normalization constant given by:
\begin{equation}
c_{\beta}=\frac{\sqrt{\pi}}{Z_{4,\beta}^{\mathrm{(H)}}}8^{3\beta+1}\Gamma\left( \frac{3}{2}+3\beta\right).
\end{equation} 
The function $P_{\beta}(r)$ is symmetric or antisymmetric depending on the value of $\beta$. This can be shown as follows. If $r$ is expressed as $r=(\lambda_{4}-\lambda_{2})/(\lambda_{3}-\lambda_{1})$, one obtains $-r$ upon exchanging $\lambda_{4}$ and $\lambda_{2}$.  Spacings $s_{k}=\lambda_{k+1}-\lambda_{k}$ then transform as
\begin{align}
s_{1}\rightarrow s_{1}^{\prime} &=s_{1}+s_{2}+s_{3}, \label{change_of_var_spacing1}\\
s_{2}\rightarrow s_{2}^{\prime} &= -s_{3},\\
s_{3} \rightarrow s_{3}^{\prime} &=-s_{2}\label{change_of_var_spacing3}.
\end{align}

We introduce as before a dimensionless variable $z^{\prime}=s_{1}^{\prime}/s_{3}^{\prime}$.
Using \eqref{change_of_var_spacing1}-\eqref{change_of_var_spacing3} and the fact that
\begin{equation}
s_{2}=\frac{s_1r-s_3}{1-r},
\end{equation}
one finally gets the expressions 
\begin{equation}
z=\frac{z^{\prime}+r}{1+rz^{\prime}}, \quad z^{\prime}=\frac{z-r}{1-rz},
\end{equation}
where $z=s_1/s_3$ as before. The denominator $G(r,z)=\left[ h(r)(1+z^2)-2(1+r)^2z\right]^{3/2+3\beta}$ appearing in the integrand satisfies a functional equation. Namely, for all $u$,
\begin{equation}
G\left(-r,\frac{u-r}{1-ru}\right)=\left(\frac{1+r}{1-ru}\right)^{3+6\beta} G(r,u).
\end{equation}
Changing $r$ to $-r$ in \eqref{result_final_hermite_rgrand} and then performing the change of variable $z=(u-r)/(1-ru)$, one easily verifies that
\begin{equation}
P_{\beta}(-r)=(-1)^{\beta+1}P_{\beta}(r). \label{parity_distribution}
\end{equation}
Performing the integral in \eqref{result_final_hermite_rgrand}, one obtains
\begin{equation}
P_{\beta}(r)=(-1)^{\beta+1}\Lambda^{(\beta)}(r)+\Lambda^{(\beta)}(-r), \label{Result_final_overlap_hermite}
\end{equation}
with 
\begin{equation}
\Lambda^{(\beta)}(r)= d_{\beta}\frac{r^{4\beta}}{(1+r^{2})^{3\beta+1}}\frac{|r-1|^{2\beta+1}}{h(r)^{1/2+2\beta}}Q^{(\beta)}\left(r+\frac{1}{r}\right), \label{Lambda_function}
\end{equation}
where $d_{\beta}$ is a constant and $Q^{(\beta)}$ are  polynomials of degree $3\beta$. For the different $\beta$ ensembles these polynomials read:
\begin{align}
Q^{(1)}(w)&=12-36w+20w^2+15w^3, \quad d_{1}=8,\\
Q^{(2)}(w)&=1920 - 10560w + 19184w^2 - 8552w^3 - 9124w^4 + 5454w^5 + 2727w^6, \quad d_{2}=\frac{8}{\pi}, \\
\notag Q^{(4)}(w)&=3244032 - 34062336w + 146853888w^2 - 320587776w^3 + 
 322416384w^4\\
 \notag & + 12364416w^5 - 318898752w^6 + 172975392w^7 + 
 113704048w^8\\ 
 & - 96136152w^9 - 26756676w^{10} + 19539090w^{11} + 
 5861727w^{12}, \quad d_{4}=\frac{32}{3\pi}.
\end{align}
The small $r$ behavior can be obtained from the Taylor expansion of \eqref{Lambda_function} or from the integral expression and one has:
\begin{equation}
P_{\beta}(r)\sim \xi_{\beta} r^{3\beta+ 1},
\end{equation}
with 
\begin{equation}
\xi_{1}=\frac{2240}{81\sqrt{3}} \simeq 15.9662,\qquad\xi_{2}=\frac{512512}{729\pi\sqrt{3}} \simeq 129.201,\qquad\xi_{4}=\frac{174054932480}{4782969\pi\sqrt{3}}\simeq 6687.72.
\end{equation}
The large $r$ behavior can be deduced from the functional equation:
\begin{equation}
P_{\beta}\left(\frac{1}{r}\right)=r^{2}P_{\beta}(r), 
\end{equation}
and one finds the following power law:
\begin{equation}
P_{\beta}(r)\sim \xi_{\beta}r^{-3\beta-3}.
\end{equation}

\begin{figure}[!t]
\begin{center}
\includegraphics*[width=0.8\linewidth]{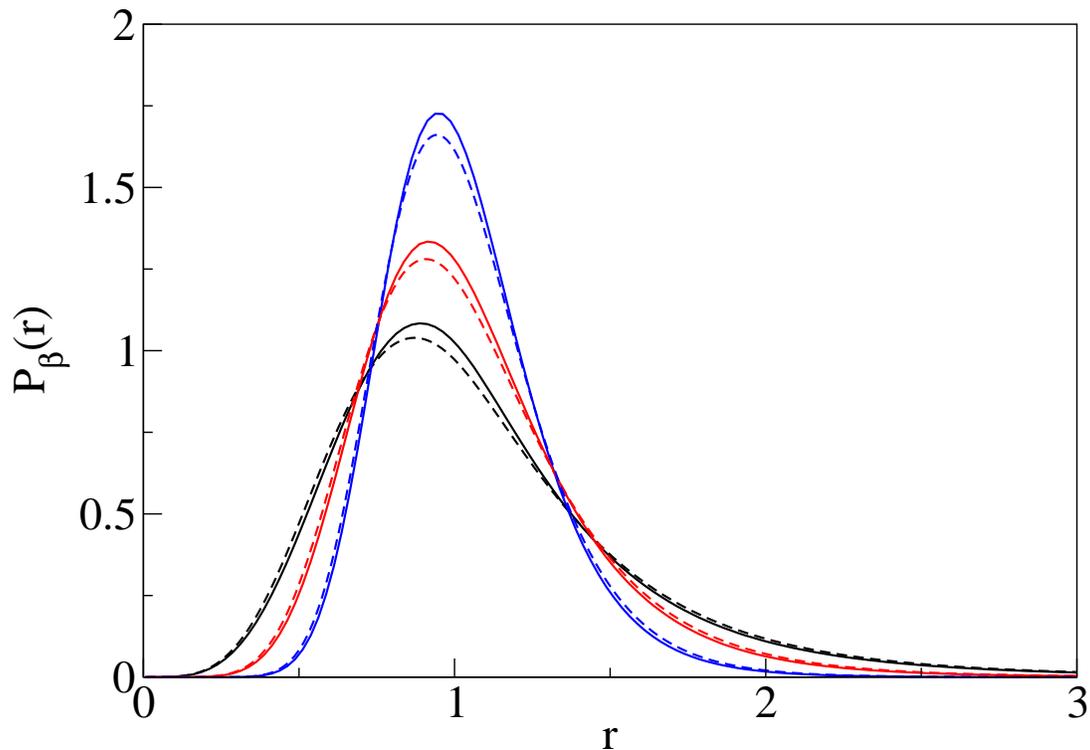}
\caption{Distribution of the overlapping ratio $r$ for the bulk eigenvalues of $10^6$ random matrix realizations of size $N=1000$ (solid). From bottom to top: GOE (black), GUE (red) and GSE (blue) obtained from tridiagonal matrices (see \cite{dumitriu}). Dashed: corresponding theoretical results \eqref{result_final_hermite_rgrand} for $N=4$. 
\label{fig.P2rbis}}
\end{center}
\end{figure}

\begin{figure}[!t]
\centering \includegraphics*[width=0.8\linewidth]{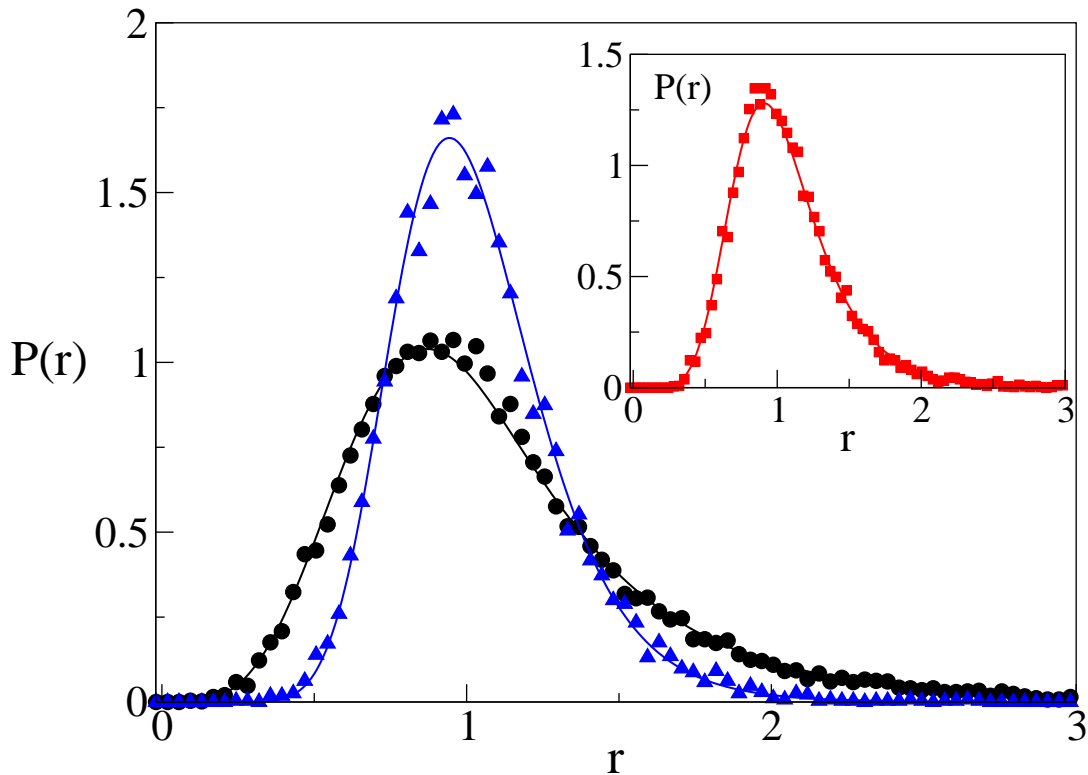}
\caption{$k=1$ overlapping distribution for the quantum Ising model  for $\lambda=0.5$, $\alpha=0.5$ and $L=18$ spins in the sector of eigenvalue $\omega_{4}$ (14588 energy levels) of the translation operator $\hat{T}$ (black circles). Blue triangles are obtained from one half of the total spectrum with every second level suppressed: we recover GSE statistics in this case as expected. Full lines correspond to prediction (\ref{Result_final_overlap_hermite}). Inset: The same for the zeros of Riemann zeta function up the critical line ($10^{4}$ levels starting from the $10^{22}$th zero, taken from \cite{odlyzko}). Full line corresponds to the expected GUE statistics (\ref{Result_final_overlap_hermite}) with $\beta=2$.\label{IsingHermite}}
\end{figure}

In Fig.~\ref{fig.P2rbis} we show how the analytical result \eqref{Result_final_overlap_hermite} compares with numerics for large size matrices. 
Clearly for such ratios the $4\times 4$ surmise is much less accurate than for the ratios \eqref{defratios}, although it reproduces the main qualitative features of the large $N$ distribution. Given the accuracy of experimental data, this already gives a useful formula, as we illustrate in Fig.~\ref{IsingHermite} on the quantum Ising model \eqref{IsingHamiltonian} and on the zeros of the Riemann zeta function.

\section{Conclusions}\label{conclusions}

In this paper, we considered various generalizations and extensions of the calculations in \cite{atas} concerning ratios of consecutive spacings between eigenvalues of random matrices. First, we derived the distribution of ratios in the case of $N=4$ eigenvalues, obtaining an explicit analytic expression that is more accurate by an order of magnitude than the expression of \cite{atas}. Next, we generalized this result by providing an exact general formula for the joint probability density $\cp_\beta^{(V)} (r_1,\ldots, r_{N-2})$ of the ratios $r_j$ of consecutive spacings, valid for any $\beta$-ensemble of random matrices characterized by the confining potential $V(x)$. The general formula \eqref{general} was then specialized to the $\beta$-Hermite (formula \eqref{finalgaussian}) and $\beta$-Laguerre (formula \eqref{finalwishart}) cases. The surmise \eqref{surmise3} then becomes a special case of the general formula \eqref{finalgaussian} for $N=3$, and derived expressions in the cases $N=4,5$. In general, we observed a universal behavior $\rho_{\beta,N}^{(V)}(r)\sim r^\beta$ for small $r$ and $\rho_{\beta,N}^{(V)}(r)\sim r^{-2-\beta}$ for large $r$, in full agreement with the surmise in \cite{atas}. We also give an example of a possible explicit calculation by computing the average ratio $\langle r\rangle$ for the $\beta$-Laguerre ensemble for $N=5$. This gives an explicit algorithm for the closed-form evaluation of averages over $\hat{\cp}_\beta^{(V)} (f_1,\ldots,f_{N-2})$, at least for not too large $N$. Finally, we proposed a new useful ratio statistics, namely the $k$th overlapping ratio, for which we provided analytic expressions. For Poisson distribution we obtained an exact formula for any $k$. For $\beta$-Hermite ensembles of random matrices, we derived a formula  for $N=4$ in the case $k=1$ which describes well the asymptotic behaviour of the standard random matrix ensembles for large matrix size. Our results were then applied to spectral properties of a quantum Ising model and to zeros of the Riemann zeta function. In view of potential extension of our work, it is interesting to notice that the joint density of the auxiliary variables $f_j$ (once fully symmetrized) resembles the joint density of eigenvalues of a non-classical invariant ensemble of matrices, since the interaction is again of the Vandermonde type. It would be interesting to investigate whether the standard analytical machinery (orthogonal polynomial technique and Coulomb gas method) well-suited to the invariant ensembles could be possibly borrowed and applied to these cases as well.

 \begin{acknowledgments}
The authors thank Guillaume Roux for useful discussions. PV is grateful to Clarissa Dell'Aquila for insightful discussions, and acknowledges financial support from Labex/PALM (project Randmat). YYA was supported by the CFM foundation. 
 \end{acknowledgments}

\end{document}